# The Influence of A-site Ionic Radii on the Magnetic Structure of Charge Ordered $La_{0.5}Ca_{0.5-x}Sr_xMnO_3$ Manganites

**Indu Dhiman, A. Das, P. K. Mishra[*] and L. Panicker**

*Solid State Physics Division, Bhabha Atomic Research Centre, Mumbai - 400085, INDIA.*

[*]*Technical Physics and Prototype Engineering Division, Bhabha Atomic Research Centre, Mumbai - 400085, INDIA.*

The influence of the A-site ionic radii ($<r_A>$) on the magnetic structure has been investigated in $La_{0.5}Ca_{0.5-x}Sr_xMnO_3$ compounds ($0 \leq x \leq 0.5$) using neutron diffraction, magnetization, and resistivity techniques. All compounds in the composition range $x = 0.3$ crystallize in the orthorhombic structure (Space Group *Pnma*). No further structural transition is observed as temperature is lowered below 300K. The compound $x = 0.4$, is a mixture of two orthorhombic phases crystallizing in *Pnma* and *Fmmm* space group. The $x = 0.5$ compound has a tetragonal structure in the space group, *I4/mcm*. The charge ordered (CO) state with CE-type antiferromagnetic order remains stable for $x = 0.3$. Above $x = 0.3$, the CE-type antiferromagnetic state is suppressed. In $x = 0.4$ compound, A-type antiferromagnetic ordering is found at temperatures below 200K. Orbital ordering accompanying the spin ordering is found in all the samples with $x = 0.4$. The system becomes ferromagnetic at $x = 0.5$ and no signature of orbital ordering is observed. As a function of $<r_A>$, the charge ordered state is stable up to $<r_A> \sim 1.24$Å, and is suppressed thereafter. The magnetic structure undergoes a transformation from CE-type antiferromagnetic state to a ferromagnetic state with an intermediate A-type antiferromagnetic state.

# I. Introduction

Intense research on experimental and theoretical fronts have been focused on charge ordered manganites due to the coexistence of charge, orbital, and spin ordering at various temperatures. Studies on charge ordered manganites have shown that the charge ordered state is sensitive to the average size of the A-site cation $\langle r_A \rangle$, hydrostatic pressure, magnetic field, chemical substitutions at the Mn site, and the A-site ionic radii mismatch ($\sigma^2$) effects [1, 2]. The A-site ionic radii mismatch i. e. disorder is quantified by the variance of the A-cation radius distribution expressed as $\sigma^2 = \sum x_i r_i^2 - \langle r_A \rangle^2$, where $x_i$ denotes the fractional occupancy of the A-site ion and $r_i$ is the corresponding ionic radius. The variance $\sigma^2$ provides a measure of the oxygen displacement Q due to A-site cation disorder [3]. Charge ordered manganites in general show different types of ground states depending on the dominance of antiferromagnetic (AFM) and / or ferromagnetic (FM) interactions, and Jahn-Teller distortions. The half-doped perovskite manganites $R_{0.5}A_{0.5}MnO_3$ (R = Trivalent Rare Earth ion, A = Divalent ion Ca, Sr, Ba) exhibit a wide variety of magnetic structure and magnetotransport behavior and have been extensively investigated [4]. However, the complex physics behind these have not been fully comprehended and therefore, call for further studies. Depending on the ionic radii of R and A cations, these compounds exhibit different structural, transport and magnetic properties. The compound $La_{0.5}Ca_{0.5}MnO_3$ with average ionic radius $\langle r_A \rangle$ of 1.198Å is particularly interesting as it undergoes successive ferromagnetic metallic ($T_C \sim 230K$), antiferromagnetic insulating ($T_N \sim 170K$), charge and orbital ordering transitions. The AFM ordering in the presence of charge ordering is found to be of CE-type [5-7]. The charge ordering effect, i.e. a regular arrangement of the $Mn^{3+}$ and $Mn^{4+}$ ions, occurs below a temperature $T_{CO}$, which may coincide with $T_N$. As a result of charge ordering below $T_{CO}$, the carriers are localized into specific sites, giving rise to long range ordering in the crystal. In $La_{0.5}Ca_{0.5}MnO_3$ the spin and charge ordering transition is accompanied by orbital ordering, in which according to Goodenough's model, $d_{z^2}^2$ $Mn^{3+}$ orbitals (associated with long $Mn^{3+}$-O bond lengths in the Jahn-Teller distorted $Mn^{3+}O_6$ octahedra) would order, forming zig - zag chains in the a-c plane [5, 8]. On the other hand, the system $La_{0.5}Sr_{0.5}MnO_3$ with larger average A-site ionic radius $\langle r_A \rangle$ (1.263Å) is reported to be a ferromagnetic metal with ferromagnetic transition temperature, $T_C \sim 310K$. In contrast to $La_{0.5}Ca_{0.5}MnO_3$ this compound does not show charge or orbital ordering. However, a weak A-type antiferromagnetic ordering in addition to ferromagnetic ordering has been observed in this compound [9].

In other Sr doped compounds like $Nd_{0.5}Sr_{0.5}MnO_3$ ($<r_A>$ = 1.236Å) and $Pr_{0.5}Sr_{0.5}MnO_3$ ($<r_A>$ = 1.245Å) it is found that the former compound with smaller $<r_A>$ is a charge ordered insulator ($T_{CO}$ ~ 158K) with CE-type antiferromagnetic spin structure below 150K while the latter with larger $<r_A>$ exhibits A-type spin structure below its ordering temperature [10, 11]. An analogous change in magnetic structure as a function of $<r_A>$ has also been observed in $Pr_{0.5}Sr_{0.5-x}Ca_xMnO_3$ and $Pr_{0.5}Sr_{0.5-x}Ba_xMnO_3$ systems [12-14]. In $(Nd_{1-z}La_z)_{0.5}Sr_{0.5}MnO_3$ system, partial replacement of $Nd^{3+}$ by a larger $La^{3+}$ ion suppresses the charge ordered state. As z increases, A-type antiferromagnetic state is observed as an intermediate state [15].

Application of hydrostatic pressure on these systems is equivalent to increasing $<r_A>$. Interestingly, even though increasing $<r_A>$ or applying pressure has opposite effects on volume, they have similar effects on magnetic and transport properties. The application of hydrostatic pressure results in a transformation of the CE-type AFM structure to A-type antiferromagnetic structure in $(Nd_{1-z}La_z)_{0.5}Sr_{0.5}MnO_3$ system for z = 0.4 composition [15]. High pressure studies on charge ordered $Nd_{0.5}Sr_{0.5}MnO_3$ system reveals that on application of high pressure (≥ 4.5 GPa) A-type AFM state with orbital ordering is stabilized at the expense of CE-type AFM state [16]. In view of the existing literature on charge ordered manganites discussed above, it is evident that change in $<r_A>$ affects the magnetic structure. Thus, it would be illuminating to study the variation of magnetic structure as a function of $<r_A>$.

The results summarized above indicate that the substitution of cations of different sizes at the rare earth sites results in lattice distortions that may influence the FM double exchange and the AFM super-exchange interactions differently. Generally, for large $<r_A>$, a charge delocalized ferromagnetic state is stable at low temperatures, whereas for small $<r_A>$ the system goes to a charge ordered, AFM, insulating state. These results can be understood in terms of one electron bandwidth, $W$ of $e_g$ electrons which depends on the Mn-O-Mn bond angle according to the formula given as, $W \alpha \dfrac{Cos \frac{1}{2}(\pi - <\theta>)}{d^{3.5}}$ where, $<\theta>$ is the average Mn-O-Mn bond angle and d is the average Mn-O bond length [17]. This implies that as the Mn-O-Mn bond angle approaches 180°, $W$ increases, and the spatial overlap of Mn $e_g$ and O 2pσ orbitals increases favoring FM double exchange interactions. Decreasing the ionic radii $<r_A>$ reduces the Mn-O-Mn bond angle leading to narrower one electron bandwidth, $W$. This effectively favors an insulating state.

The one electron bandwidth alone, however, cannot predict reliably the occurrence of various magnetic phases in charge ordered systems. This is evident in the reported results of $La_{0.5-x}Y_xCa_{0.5}MnO_3$ compound. In these compounds substitution of Y leads to reduction in $<r_A>$ while $\sigma^2$ increases. This substitution results in suppression of antiferromagnetic ordering. The one electron bandwidth model however, predicts an enhancement of antiferromagnetic ordering. These observations are correlated with the effects of disorder [7, 18]. However, according to Vanitha et. al. for fixed hole concentration and $<r_A>$, it is found that charge ordered transitions are not very sensitive to the mismatch between the sizes of the A-site cations [19].

The ionic radii of $Pr_{0.5}Ca_{0.5-x}Sr_xMnO_3$ series and $La_{0.5}Ca_{0.5-x}Sr_xMnO_3$ series are similar. However, the disorder parameter $s^2$ is much reduced for La based series and is almost constant for x ranging from 0.2 to 0.5. This signifies that this series is a good candidate to study the effect of $<r_A>$ independent from the disorder.

To explain charge and orbital ordering, one must consider the intimate balance between a number of competing interactions such as Hund's coupling, Jahn-Teller distortions and coulomb interactions. The AFM state is stabilized by kinetic energy of $e_g$ electrons whose motion is restricted by $t_{2g}$ spin alignment through double exchange mechanism. The theoretical studies have indicated that JT-distortion plays a crucial role in the stability of A-type AFM state. Monte Carlo simulation studies carried out in charge ordered systems show that various magnetic ground states and their coexistence in the form of electronic phase separation state could be largely reproduced by the study of interplay between electron phonon coupling and Heisenberg coupling between localized $t_{2g}$ spins [20-24].

Previous studies on $La_{0.5}Ca_{0.5-x}Sr_xMnO_3$ system have indicated that $T_N$ shows a non-monotonous behavior with increasing Sr concentration up to x = 0.3 and finally disappears for x = 0.4 [25]. This variation in the magnetic properties as a function of Sr concentration suggests a possible change in the type of magnetic structure in the antiferromagnetic state. Neutron diffraction is a powerful tool to investigate such a change in magnetic structure.

In this paper we report the effects of varying $<r_A>$ on the evolution of magnetic structure in the half doped $La_{0.5}Ca_{0.5}MnO_3$. The substitution of smaller A-site cation with a larger ion is equivalent to applying hydrostatic pressure. It would be interesting to study whether these two effects lead to similar changes in magnetic structure. We have substituted $Ca^{2+}$ ion ($r_{ion}$ = 1.18Å) with larger $Sr^{2+}$ ions ($r_{ion}$ = 1.31Å) to obtain $La_{0.5}Ca_{0.5-x}Sr_xMnO_3$ (0.1 ≤ x ≤ 0.5) [26]. Increasing the A-site ionic size $<r_A>$ leads to increase in disorder up to x ≤ 0.4.

At x = 0.5, $\langle r_A \rangle$ increases but $s^2$ decreases due to complete substitution of Ca by Sr. In $La_{0.5}Ca_{0.5}MnO_3$ compound, the presence of FM and AFM-CO phases at different temperatures are advantageous for studying the effect of $\langle r_A \rangle$ and $s^2$ on both these phases. The aim of our study is to investigate the changes in magnetic structure arising as a consequence of increase in $\langle r_A \rangle$ and therefore, study the transition from an inhomogeneous AFM state (x = 0) to a homogeneous ferromagnetic state (x = 0.5). We induce size disorder by substitution of $Ca^{2+}$ ion by $Sr^{2+}$ ion into the A cation site. Sr substitution for Ca is carried out keeping $Mn^{3+}$ to $Mn^{4+}$ ratio constant. Therefore, all the observations reported are a direct consequence of average A-site cationic radii $\langle r_A \rangle$ and cation size disorder $s^2$.

## II. Experiment

The samples were synthesized by conventional solid-state reaction method. The starting materials $La_2O_3$, $MnO_2$, $SrCO_3$, and $CaCO_3$ were mixed in stoichiometric ratio and kept for calcination at 1150ºC for 24hrs. Samples were then removed and kept for sintering at 1400ºC for 24hrs with intermediate grinding. Finally, the samples were pelletized and heat treated at 1400ºC for 24hrs.

The phase purity of the final products was ensured by X-ray powder Diffraction at 300K with a Rigaku diffractometer, rotating anode type using Cu K$\alpha$ radiation of wavelength, $\lambda = 1.544$Å between $10º \leq 2\theta \leq 70º$. All compositions reported here are found to be single phase except x = 0.1 and 0.2 in which, a small amount of $Mn_3O_4$ impurity phase was present. These were accounted for in neutron diffraction studies.

Neutron diffraction patterns were recorded on a multi PSD based powder diffractometer, ($\lambda = 1.249$Å) at Dhruva reactor, BARC, Mumbai at selected temperatures between 17K and 300K in the $5º \leq 2\theta \leq 140º$ range. The powdered samples were packed in a cylindrical Vanadium container and attached to the cold finger of a closed cycle Helium refrigerator. Rietveld refinement of the neutron diffraction patterns were carried out using FULLPROF program [27]. The magnetization measurements were carried out on a Superconducting Quantum Interference Design (SQUID) magnetometer. The zero field (ZFC) and field cooled (FC) measurements were performed under a magnetic field of 1Tesla. The dc resistivity measurements between 3K and 300K were performed by standard four probe technique. Differential Scanning Calorimetry (DSC) measurements were carried out using DSC-822 METTLER TOLEDO in the temperature range 123K to 423K with scanning rate of 10K/min. For this measurement the samples were sealed in an Al pan, with an empty pan as reference. The instrument was calibrated using high purity Indium and Cyclohexane for measuring temperature and enthalpy.

## III. Results and Discussion

### A. Crystal Structure

The series $La_{0.5}Ca_{0.5-x}Sr_xMnO_3$ for $x \leq 0.3$ crystallizes in single-phase orthorhombic structure (space group *Pnma*), as observed in the case of parent compound $x = 0$. This observation is in contradiction with the reported monoclinic structure (Space Group *I2/a*) in samples $La_{0.5}Ca_{0.5-x}Sr_xMnO_3$ ($0.2 \leq x \leq 0.4$) at room temperature [25]. In figure 1 we have shown the neutron diffraction pattern in $2\theta$ range 55° to 60° for the compound $x = 0.3$ fitted in *I2/a* and *Pnma* space group separately. It can be observed that *I2/a* space group ($\chi^2 = 3.29$, $R_{Bragg} = 14.9$, and $R_f = 13.1$) does not fit the observations as well as the *Pnma* space group ($\chi^2 = 1.66$, $R_{Bragg} = 6.95$, and $R_f = 6.04$). On lowering of temperature possibility of a structural transition has been reported in charge ordered compounds $x = 0$ [5, 28]. Lowering of symmetry to $P2_1/n$ space group is used to accommodate the ordering of $Mn^{3+}$ and $Mn^{4+}$ ions on to two inequivalent sites instead of a single site available in the space group *Pnma*. However, lowering of space group to $P2_1/n$ necessarily requires inclusion of 29 positional parameters as against 7 positional parameters in the case of *Pnma*. In this case the refinement becomes unstable and the obtained parameters are less reliable in the absence of a good quality synchrotron data. Such a conclusion in the case of $x = 0$ has been arrived at from analysis of synchrotron data though the neutron data in the charge ordered state was still described in the *Pnma* space group [5]. We find that refining our data in $P2_1/n$ space group does not lead to appreciable improvement in $\chi^2$. Therefore, we have analyzed the neutron diffraction data in *Pnma* space group, though we agree this gives an average picture and possibility of a transition to lower symmetry exists in these compounds of which we are unable to comment from this data. At composition $x = 0.4$, sample is found to be a mixture of two orthorhombic phases of space groups *Fmmm* and *Pnma*. The volume fractions of the two phases exhibit temperature dependence as shown in figure 2. It is observed that at 15K the structure is consistent with 55% of orthorhombic phase with space group *Pnma* and 45% of orthorhombic phase with space group *Fmmm*. As temperature increases, *Fmmm* phase fraction increases at the expense of *Pnma* phase fraction. As is evident from figure 2, the phase fraction changes sharply around 200K. This temperature is found to coincide with the antiferromagnetic transition temperature obtained from neutron diffraction data. Outside the 150 – 225K window, where phase fraction change dramatically, there is no notable change in the phase fraction. The end composition i. e., La$_{0.5}$Sr$_{0.5}$MnO$_3$, is a single-phase compound with tetragonal structure (space group *I4/mcm*). In literature, two different models have been proposed for this sample. The first one is a mixture of two phases namely, orthorhombic (*Imma*) and tetragonal (*I4/mcm*) at 300K. According to the other model the sample is a mixture of two phases in space group *Fmmm* and *I4/mcm* [29, 9]. Hence, we refined the data using the two-phase model described above. In each case, neither the $\chi^2$ of the fit nor the goodness of the fit showed any improvement over the fit obtained using both two-phase models mentioned above. Therefore, we conclude that the sample La$_{0.5}$Sr$_{0.5}$MnO$_3$ is a single-phase compound with tetragonal structure (*I4/mcm*).

The unit cell volume increases with Sr$^{2+}$ doping. This is attributed to larger ionic radii of Sr$^{2+}$ in comparison to Ca$^{2+}$ (~ 11%). The variance has been estimated to vary between $0.30 \times 10^{-3}$ to $2.51 \times 10^{-3}$ for x varying between 0.1 and 0.4, respectively. The positive correlation between $\sigma^2$ and x breaks down at x = 0.5. Even though <r$_A$> increases to 1.263Å, $\sigma^2$ decreases to $2.208 \times 10^{-3}$, due to complete substitution of Ca$^{2+}$ by Sr$^{2+}$ reducing the ionic size mismatch [26]. The increase in $\sigma^2$, however, is not as large as reported in the case of Ba$^{2+}$ doped La$_{0.5}$Ca$_{0.5-x}$Ba$_x$MnO$_3$ compounds. In this series for x = 0.3, increase in variance is so large that chemical phase separation occurs driven by strain field associated with the local displacement of Oxygen ions leading to the two phase nature, observed for x = 0.4 [30]. The refined structural parameters, bond angles and lengths at 300K and 17K are summarized in Table I and Table II, respectively.

The temperature dependence of cell parameters for x = 0.3 (representative of samples with $0 \leq x \leq 0.4$) and x = 0.5 are shown in figure 3(a) and 3(b), respectively. On lowering temperature, lattice parameters exhibit anomalous behavior for samples between $0.1 \leq x \leq 0.4$, as observed in the case of parent compound. The (2 0 2) and (0 4 0) reflections which were merged at 300K exhibit a splitting, coinciding with the ordering temperature. The refined lattice parameter b shrinks drastically, while a and c parameters expand. This behavior is observed normally in systems exhibiting CE-type antiferromagnetic ordering and is associated with orbital ordering of the $d_{z^2}$ orbital in the a-c plane [5, 6]. This behavior is retained for Sr$^{2+}$ doping up to x ≤ 0.4. However, at x = 0.4 the orbital ordering is characterized by ordering of $d_{x^2-y^2}$ type orbitals within the subsequent planes [2]. This change in nature of the orbital ordering is associated with the change in magnetic spin structure from CE-type to A-type discussed later. Volume decreases continuously on lowering of temperature and no anomalous behavior around $T_C$ is observed except for a small change of slope around $T_C$. For x = 0.5 cell parameters do not show any such anomalous behavior indicating absence of orbital ordering. Similar anomalous behavior in lattice parameters attributed to orbital ordering have been also observed in other charge ordered systems like for $Nd_{0.5}Sr_{0.5}MnO_3$ ($<r_A>$ = 1.189Å) and $Pr_{0.5}Ca_{0.5-x}Sr_xMnO_3$ [31, 13].

The temperature variation of the bond length, as shown in figure 3(c) for sample x = 0.3 are similar to those of cell parameters and indicate a coupling of the static Jahn-Teller distortion of the $Mn^{3+}$ ions with magnetic ordering in this compound. Thus, two shorter $Mn-O_{ap}$ bond distances i.e. along b-axis and four longer $Mn-O_{eq}$ bond distances i.e. in a-c plane characterize the octahedra. The difference in bond lengths is maximum at low temperatures implying maximum distortion, and above ordering temperature the difference in Mn-O bond lengths is drastically reduced implying that octahedra are almost undistorted, with six almost equal Mn-O distances. The average <Mn-O> bond length remains almost constant, and no change is observed across the ferromagnetic transition temperature, as shown in figure 3(c). There are two characteristic distortions that influence the perovskite structure in manganites. The first consists of a cooperative tilting of the $MnO_6$ octahedra as a consequence of ionic mismatch. The second kind of distortion is connected with a static Jahn-Teller (JT) leading to distortion of $MnO_6$ octahedra, leading to unequal splitting of Mn-O bond lengths. In comparison, the Mn-O-Mn bond angle does not exhibit any variation with temperature. This is shown in figure 3(d). The absence of variation in the average bond length and bond angle with temperature indicates their lack of influence in the resistivity behavior. However, increasing the ionic radii $<r_A>$ leads to increase of the <Mn-O-Mn> bond angles in the basal plane from 161º (x=0.1) to 166º (x=0.4), moving closer towards ideal 180º (Table I and II). This leads to wider one electron bandwidth, *W*. Therefore the spatial overlap of Mn $e_g$ and O 2pσ orbitals increases which favors FM double exchange interactions leading to suppression of the charge ordered behavior.

### B. Magnetic and Electrical properties

The variation of magnetization as a function of temperature for samples x = 0 - 0.5 in a field of H = 1T are shown in figure 4. The magnetic field of 1T was chosen since it is above the anisotropy field and much below the values that would cause a pronounced decrease in $T_{CO}$ due to melting of the charge ordered state [32]. Magnetization data for parent compound is taken from Ref. 7 for comparison. It is reported that $La_{0.5}Ca_{0.5}MnO_3$ exhibits double transition at $T_C \sim 230K$ and $T_N \sim 170K$ [5, 6]. With Sr doping no major change in magnetization behavior is observed in comparison to the parent compound. However, the maximum value of magnetization (M) is reduced with the increase of Sr doping (from x = 0.0 to 0.3), which could be due to reduction of FM clusters size in comparison to parent compound. It is also observed that the magnitude of M in the antiferromagnetic region is almost equal to the value in the paramagnetic region. This is an indication of reduction of ferromagnetic clusters in antiferromagnetic region in contrast to the case observed in the parent compound. In the parent compound the phase below Nèel temperature is not purely antiferromagnetic, it is rather an inhomogeneous mixture of ferromagnetic and antiferromagnetic clusters. Electron microscopy experiments carried out at 90K on $La_{0.5}Ca_{0.5}MnO_3$, reveals an inhomogeneous mixture of ferromagnetic and antiferromagnetic regions [33]. For x = 0.4, the maximum value of magnetization is higher in comparison to samples with $x \leq 0.3$, although their nature of temperature dependence is similar. The enhanced magnetization indicates a higher volume fraction of the ferromagnetic phase in this compound. The plot of magnetization as a function of temperature exhibits only the ferromagnetic nature of the sample, x = 0.5. However, low temperature neutron diffraction pattern reveals the presence of weak antiferromagnetic reflections in addition to ferromagnetic contribution, as discussed later. The absence of signature of AFM ordering in M(T) plot is due to the dominating influence of ferromagnetic interactions over the antiferromagnetic contributions. The ferromagnetic transition temperatures, $T_C$ were obtained from the dM/dT versus T plot. It is observed that the $T_C$ increases monotonically from 244K (x = 0.1) to 277K (x = 0.3) with increasing $<r_A>$ and is in agreement with previously reported magnetization studies on similar compounds [25, 34]. The variation of $T_C$ with $<r_A>$ suggests increase in FM interactions with increase in $<r_A>$. Our measurements indicate that the end composition x = 0.5 has $T_C$ above 300K and $T_N \sim 125K$ (obtained from neutron diffraction studies). This is in agreement, albeit with different ordering temperature, with the reported magnetization studies on polycrystalline samples of $La_{0.5}Sr_{0.5}MnO_3$ which show existence of two transitions, corresponding to $T_C$ and $T_N$ of 280K and 150K, respectively [35].

The temperature dependence of the normalized resistance, R(T)/R(300K) is shown in figure 5. The charge ordered temperature was found from the minima of the d(lnR)/dT versus temperature plots. The sample x = 0.3, displays an insulating behavior as the temperature is lowered and shows a steep rise in resistivity at the CO transition temperature ($T_{CO}$ ~ 240K), which almost coincides with the peak in DSC endotherm in this compound. This behavior is similar to parent compound which, exhibits an insulating behavior with very sharp increase in the resistivity at the CO transition (~ 155K) [36]. We observe that $T_{CO} > T_N$ ($T_N$ ~ 75K) in this particular sample. Similar observation of $T_{CO} > T_N$ has been observed previously in $La_{0.5}Sr_{1.5}MnO_4$ [37] and $Pr_{0.5}Ca_{0.5}MnO_3$ compounds [38]. Further substitution with Sr results in a gradual reduction of resistance showing transformation of CO insulating state to metallic state, beyond x = 0.3. At composition x = 0.4 ($T_{CO}$ ~ 210K), the resistance decreases slightly in comparison to x = 0.3. With further increase in x the resistance behavior changes drastically indicating the suppression of CO nature. The end member i. e. $La_{0.5}Sr_{0.5}MnO_3$ shows an insulating behavior, but the order of magnitude of change in resistance is much lower than compound with x = 0.3. Above 275K it exhibits metal like behavior. This steep decline in resistance with very little change in composition indicate the first order nature of transition from the charge ordered insulating to ferromagnetic metal like behavior. In half doped manganites the antiferromagnetic charge ordered insulating phase competes with ferromagnetic metallic phase due to double exchange mechanism. The transfer integral of the $e_g$ carriers is expressed as $t \propto t_o \cos(\theta/2)$, where $t_o$ is the transfer integral and $\theta$ is the angle between neighboring $t_{2g}$ spins. The substitution of Ca by Sr leads to enhancement in <Mn-O-Mn> bond angle (Table I and II). This leads to an increase of the transfer integral in the double exchange model, thus implying that substituting the smaller ion (Ca) by larger ion (Sr) give rise to ferromagnetic metallic state. This suggests that increase in $<r_A>$ leads to increase in one electron bandwidth, *W* and consequently the double exchange ferromagnetic interactions.

### C. Magnetic Structure and Phase Diagram

Neutron diffraction pattern of the compound $La_{0.5}Ca_{0.2}Sr_{0.3}MnO_3$ (x = 0.3) is shown in figure 6. This figure is a representative for all samples with x ≤ 0.3. In these samples AFM superlattice reflections indexed as (0 1 ½) and (½ 1 ½), characteristic of a CE-type antiferromagnetic ordering are observed below the antiferromagnetic transition temperature ($T_N$). These superlattice reflections are indexed on a 2a × b × 2c cell, having CE-type structure, which is characterized by two different Mn sublattices as proposed by Wollen and Kohler for the parent compound [39]. The diffraction pattern could be fitted to the CE-type AFM structure in *P2₁/m* space group as reported for the parent compound [5, 7]. In this structure Mn occupies distinct sites for $Mn^{3+}$ and $Mn^{4+}$. The $Mn^{3+}$ sublattice is associated with a propagation vector (0 0 ½) and $Mn^{4+}$ with (½ 0 ½). On refining the diffraction pattern for x = 0.3 the magnetic moment obtained on $Mn^{3+}$ and $Mn^{4+}$ sites at 17K are ~ 1.4$\mu_B$/Mn. The magnetic moment obtained are same on the two sites $Mn^{3+}$ and $Mn^{4+}$ indicating the possibility that the oxidation state of these two sites are nearly equivalent. A recent x-ray resonant scattering in CO systems show presence of fractional charge segregation. The oxidation states of Mn in CO state appear to differ by 0.2 electron and not 1 electron as expected [40]. The values of $T_N$ and refined magnetic moments for all the compositions at 300K and 17K are given in table I and table II, respectively. The inset in figure 6 shows the variation of integrated intensity of the AFM superlattice reflection (½ 1 ½) with temperature for $La_{0.5}Ca_{0.2}Sr_{0.3}MnO_3$ (x = 0.3). It illustrates the absence of antiferromagnetic ordering above transition temperature ~ 100K. Additionally, no ferromagnetic contribution is found in the fundamental Bragg reflections, which is in agreement with the magnetization studies (fig 4). For x= 0.4 the antiferromagnetic structure changes from CE-type to A-type. This can be observed in figure 7 where the A-type antiferromagnetic superlattice reflections are visible. The A-type AFM reflections are indexed on a × b × 2c cell in *Fmmm* space group. The refined antiferromagnetic moment at 17K is 2.86 $\mu_B$/Mn and is oriented in the *ab* plane. The temperature dependence of the refined magnetic moment for antiferromagnetic phase is shown in the inset (a) of figure 7. In this sample, in addition to superlattice reflections, we find a weak enhancement in intensity of the low angle fundamental nuclear reflections, (1 0 1) (0 2 0). Temperature dependence of integrated intensity of (1 0 1) (0 2 0) reflection is shown in the inset (b) of figure 7. This enhancement in intensity of low angle nuclear reflections is visible only close to the transition temperature, $T_C$. This is a signature of the presence of ferromagnetic ordering above the antiferromagnetic ordering temperature. This behavior is similar to enhancement in magnetization observed in the plot of M(T) as shown in figure 4. At x = 0.5, the sample exhibits predominantly ferromagnetic behavior as shown in the diffraction pattern at 17K and 300K in figure 8. The ferromagnetic phase is fitted in tetragonal structure in space group *I4/mcm*. The refined FM magnetic moment at 17K is 2.80$\mu_B$/Mn and is oriented along the c-axis. The value of the magnetic moment is in reasonable agreement with the expected FM moment of 3.5$\mu_B$/Mn. The sample x = 0.5, exhibits signature of ferromagnetism at all temperature up to 300K. Therefore, we conclude that $T_C$ for x = 0.5 is greater than 300K, in agreement with $T_C$ ~ 310K as reported previously [9]. In addition to the ferromagnetic contribution, A-type AFM superlattice reflections of very weak intensity is observed below the antiferromagnetic ordering temperature, $T_N$ ~ 125K, which is in agreement with previously reported studies on similar compound [9]. This superlattice reflection could be indexed on a × a × 2c cell in *I4/mcm* space group. The temperature dependence of the refined magnetic moment for antiferromagnetic and ferromagnetic phase is shown in inset of figure 8. However, the antiferromagnetic moment at 17K is ~ 0.6 $\mu_B$/Mn which is very small as compared to the full expected moment. This indicates that the volume fraction of the AFM phase is very low as compared to the ferromagnetic phase. This antiferromagnetic state could be a result of oxygen deficiencies observed in this sample. Therefore, it is observed that as $<r_A>$ increases the present system undergoes a transition from CE-type antiferromagnetic state to a fully ferromagnetic state with an intermediate A-type antiferromagnetic state. It is of interest to know that similar behavior is observed on increasing $<r_A>$ by partial replacement of $Nd^{3+}$ by a larger $La^{3+}$ ion, in $(Nd_{1-z}La_z)_{0.5}Sr_{0.5}MnO_3$ system [15]. The substitution of Ca with Sr widens the one electron bandwidth (*W*), leading to strong enhancement of the itinerant character of $e_g$ electrons. This favors ferromagnetic metallic behavior, due to which it has been considered that such change of *W*, can be the origin of A-type AFM metallic state. The CE-type CO state is realized for small *W*. With an increase in *W*, charge ordered state is suppressed and ferromagnetic metallic state becomes prevalent. However, before the establishment of ferromagnetic metallic state, the system goes through A-type antiferromagnetic phase as observed in the present case. The ac susceptibility measurements reported on the series $La_{0.5}Ca_{0.5-x}Sr_xMnO_3$ indicate that at x ≥ 0.4 the antiferromagnetic ordering is destroyed and only the ferromagnetic behavior is retained [25]. This inference however, is in contrast to our neutron diffraction observations. We observe the presence of antiferromagnetic ordering in addition to the ferromagnetism in the samples for x ≥ 0.4 below the antiferromagnetic transition temperature. Reiterating what was mentioned earlier, an increase in $<r_A>$ causes an increase in one electron bandwidth. This would favor ferromagnetic ordering, causing an increase in $T_C$ and a reduction $T_N$. In our samples however, the variation in $T_N$ displays non monotonous behavior with Sr doping. At x = 0.4, the change in antiferromagnetic structure from CE to A-type is accompanied by an enhancement in $T_N$ to ~ 200K. Upon further addition of Sr the $T_N$ reduces to ~ 125K. According to the previously reported data on single crystals for x = 0.5 it is found that this compound has ferromagnetic transition temperature, $T_C$ ~ 360K, while no antiferromagnetic ordering is encountered [41]. Although previous powder neutron diffraction studies on this compound at 2K have shown that the ferromagnetic *I4/mcm* phase transforms partially to an A-type antiferromagnetic phase of the orthorhombic *Fmmm* symmetry with a crystallographic cell $2a_p \times 2a_p \times 2a_p$ [9]. In our samples with x = 0.5, the magnetic structure transformation is in agreement with this report, although the chemical structure retains its orthorhombic structure (Space group *I4/mcm*) down to 15K.

In comparison, in $Pr_{0.5}Sr_{0.5-x}Ba_xMnO_3$ series (0.0 ≤ x ≤ 0.2), with decreasing temperature, two magnetic states are observed FM with *I4/mcm* symmetry and AFM with *Fmmm* symmetry. In these samples, with increasing x, $T_C$ decreases, whereas $T_N$ remains nearly constant. This evolution has been attributed to the increase of variance $s^2$ from $4.29 \times 10^{-3}$ (x = 0.0) to $21.1 \times 10^{-3}$ (x = 0.5), which counterbalances the increase of $<r_A>$, generally favoring ferromagnetism [14]. In the above case, the large increase in variance causes mismatch effects to dominate ionic size effects, in contrast to the series $La_{0.5}Ca_{0.5-x}Sr_xMnO_3$. We reason that the increase in $T_C$ and decrease in $T_N$ is due to increase in A-site ionic radii for x ≤ 0.3. However, for x > 0.3 due to change in magnetic structure this monotonous behavior of $T_C$ and $T_N$ is broken.

It is interesting to note that reports on the Ba doped compounds reveal that Ba doping of x = 0.1 is sufficient to suppress the antiferromagnetism [42], whereas in Sr doped compounds the CE-type antiferromagnetism is suppressed at x = 0.4. However, recent reports on Ba doped compounds are in disagreement with our results, where the studies on $La_{0.5}Ca_{0.5-x}Ba_xMnO_3$, revealed that increasing $<r_A>$ leads to stabilization of CO-AFM state as against the expected FM state. In this series it appears that the localizing effects of A-site cation disorder compensates for the charge delocalization induced by the increase of $<r_A>$ [30].

Figure 9 shows the differential scanning calorimetry (DSC) data for the samples 0 = x = 0.5. The plots clearly show the endothermic transitions. The temperatures and the change in entropy of these peaks are given table III. In DSC the integrated area of the endothermic peak gives the enthalpy change accompanying the transition. Besides the magnetic degrees of freedom, the heat transfer also account for the entropy and lattice energy gain due to the electronic delocalization. Therefore, local lattice structural changes arising from the delocalization of the polaronic charge carriers are also contributing to the observed endothermic peaks [43]. The DSC curves represent the first order endothermic phase transitions. The endothermic peak temperatures in DSC plot appears close to peaks in temperature variation of magnetization curves, shown by arrow in M(T) plot in figure 4. This possibly indicates that the ferromagnetic to antiferromagnetic transitions observed in samples are of first order type. With $Sr^{2+}$ doping for x = 0.3 the endothermic phase transition temperature increases. At x = 0.4, the transition temperature reduces, while it increases to 313K for x = 0.5. The change in Gibbs free energy in these transition are expressed as $\Delta G = \Delta H - T\Delta S$. When transition occurs, the system reaches an equilibrium state implying $\Delta G = 0$. Therefore, $\Delta S = \Delta H/T$. Generally the observed endothermic transitions ($\Delta H > 0$) indicate that the disorder in the system increases. The gain in total entropy can be calculated from the above formula. The total entropy of the system increases as the transition temperature of the system increases with $Sr^{2+}$ doping for x ≤ 0.4, while at x = 0.5 it reduces. It may be deduced that as x increases the disorder in the system increases, while at x = 0.5 it reduces, due to Ca being completely replaced by Sr. In the DSC studies reported for parent compound, no endothermic transition peak is observed. In the parent compound, it is assumed that interaction between lattice and polarons is attributed to hole concentration and the difference of ionic radii between $La^{3+}$ and $Ca^{2+}$ in addition to $Mn^{3+}$ and $Mn^{4+}$. These effects cancel each other in $La_{0.5}Ca_{0.5}MnO_3$ and as a result no transition is observed in the measured temperature range [44]. However, this result is in contrast with our observation for x = 0 where, an endothermic peak at 206K is observed. However, the change in entropy is very small.

Figure 10 shows a phase diagram for the series $La_{0.5}Ca_{0.5-x}Sr_xMnO_3$ from the results obtained from present and previously reported studies. The ferromagnetic Curie temperature increases continuously with increasing Sr content up to x ≤ 0.3 due to increase in $<r_A>$. The charge ordered CE-type AFM states exists for samples within the range from x = 0.0 to x = 0.3. When the Sr concentration is further increased (x ≥ 0.3), the charge ordered state slowly vanishes due to the competing double exchange and super exchange interactions, the magnetic spin structure changes from CE-type to A-type. It is possible that the $<r_A>$ value of 1.24Å defines a limit at which magnetic and transport properties drastically change, as observed in our studied samples. This feature is in agreement with earlier reported studies on $La_{0.5}Ca_{0.5-x}Ba_xMnO_3$, $Ln_{0.5}Ca_{0.5}MnO_3$ (Ln = Nd, Sm, Gd, Dy, Y) and $Nd_{0.5-x}Pr_xSr_{0.5}MnO_3$ [42, 45, 46].

Based on Monte Carlo simulation studies a phase diagram has been proposed for charge ordered Manganites [21]. These studies show that the charge ordered system exhibit CE-type and A-type antiferromagnetic as well as ferromagnetic metallic states together with orbital ordering depending up on electron phonon coupling ($\lambda$) and the coupling between the $t_{2g}$ spins ($J_{AF}$). Our results are in qualitative agreement with the proposed phase diagram in the weak electron phonon coupling limit. In this limit the system is a ferromagnet for low $J_{AF}$. With further increase in $J_{AF}$ it exhibits a transition from ferromagnetic state to CE-type antiferromagnetic state. The two regions are separated by a first order transition as is evident from our resistivity experiments. However, we also find that an intermediate magnetic phase with A-type ordering exists between the ferromagnetic and CE-type antiferromagnetic phase. A recent theoretical study on the influence of disorder in CO manganites reproduces the CE-type, A-type, and FM states with increase in $\lambda$ [24]. Therefore, the effects of change in ionic radii could be related to the tuning of the parameter $J_{AF}$ in this model. Further experiments are planned to shed more light on this issue.

## IV. Conclusions

We have studied the effect of ionic size on the magnetic structure by varying $<r_A>$ and $\sigma^2$ in the half doped system $La_{0.5}Ca_{0.5-x}Sr_xMnO_3$. Substitution of Ca by Sr leads to suppression of the charge ordered state, beyond x = 0.3. This is attributed to increase in one electron bandwidth, *W*. Our neutron diffraction measurements indicate that the suppression of the charge ordered state is accompanied by decline of CE-type antiferromagnetic ordering. An A-type antiferromagnetic ordering is observed in case of x = 0.4 sample. The emergence of FM ordering on increase of $<r_A>$ is found to disrupt the charge and orbital ordering in these compounds. The substitution of Ca with Sr widens the one electron bandwidth (*W*), leading to strong enhancement of the itinerant character of $e_g$ electrons. This favors ferromagnetic metallic state; such a change of *W* can be the origin of A-type AFM metallic state. However, the one electron bandwidth model may not be sufficient to explain the occurrence of different magnetic structures. The magnetic structures observed on changing $<r_A>$ are identified with Monte Carlo simulation studies on manganites which reveal that electron phonon coupling ($\lambda$) and antiferromagnetic coupling ($J_{AF}$) between nearest neighbor $t_{2g}$ spins play a crucial role in stabilizing these magnetic structures.

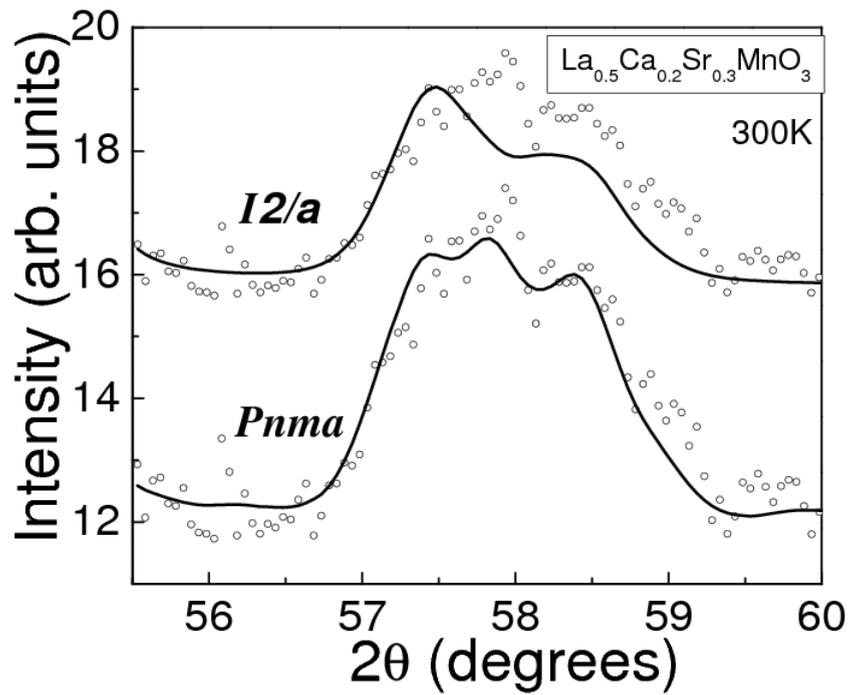

Figure: 1. A selected region (55.5 – 60°) of the fitted neutron powder diffraction patterns in two different space groups for $La_{0.5}Ca_{0.2}Sr_{0.3}MnO_3$ sample at 300K. The observed data points are indicated with open circles while the calculated pattern is shown as a continuous line.

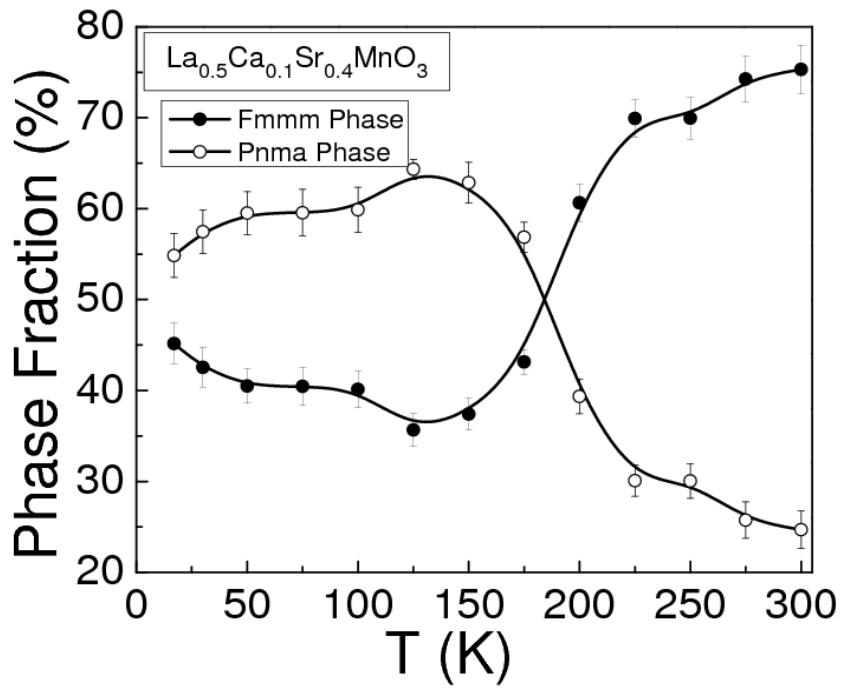

Figure: 2. Volume fraction of the *Fmmm* and *Pnma* phase with respect to temperature of the sample $La_{0.5}Ca_{0.1}Sr_{0.4}MnO_3$ is shown. The continuous lines are a guide for the eye.

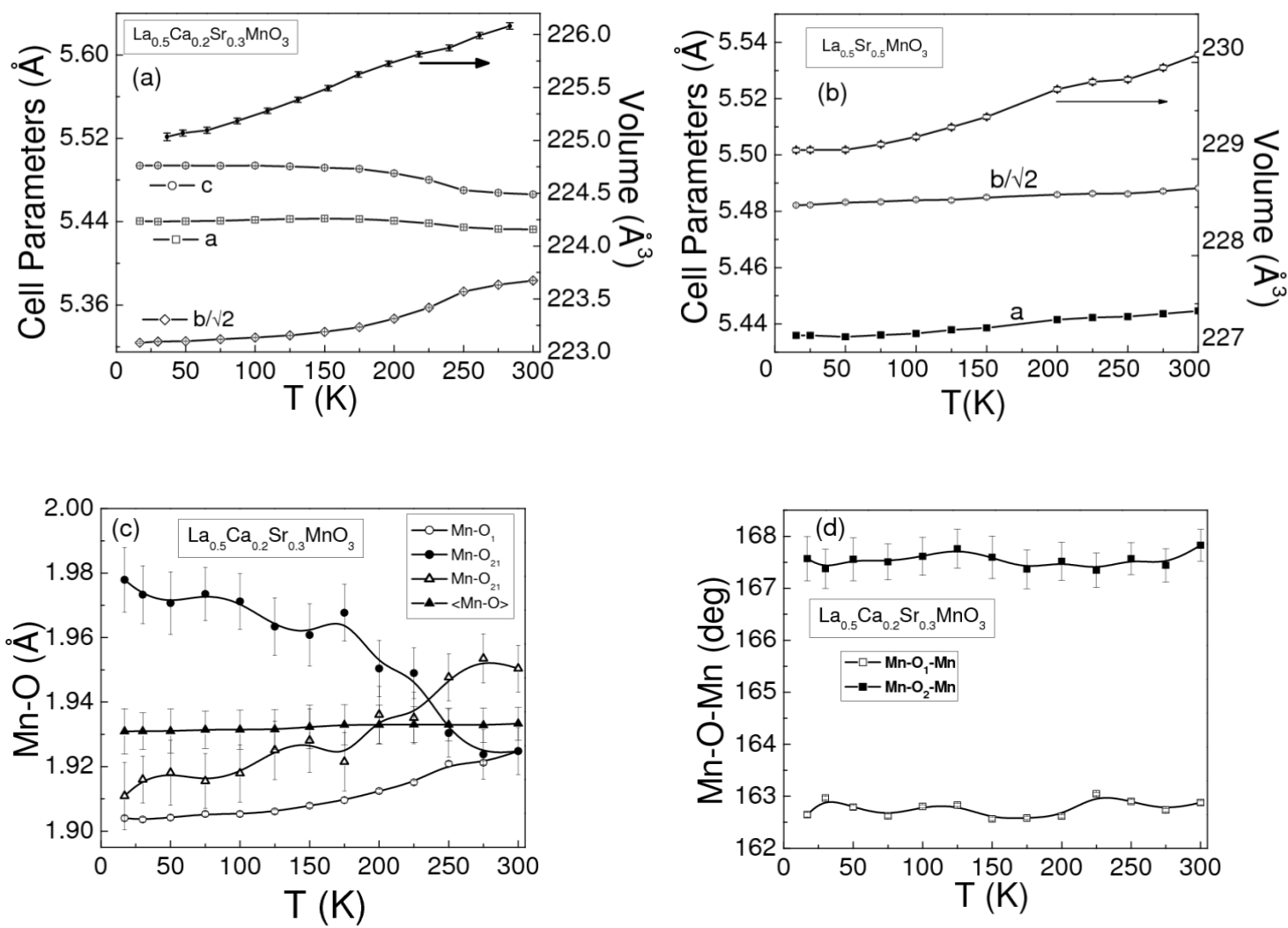

Figure: 3. (a) Temperature dependence of lattice parameters and the unit-cell volume for $La_{0.5}Ca_{0.2}Sr_{0.3}MnO_3$ sample (b) Variation of lattice parameters for sample $La_{0.5}Sr_{0.5}MnO_3$ and unit-cell volume with temperature (c) The variation of Mn-O bond distances with temperature for $La_{0.5}Ca_{0.2}Sr_{0.3}MnO_3$ sample and (d) Temperature dependence of Mn-O-Mn bond angles for $La_{0.5}Ca_{0.2}Sr_{0.3}MnO_3$ sample. The continuous lines are a guide for the eye.

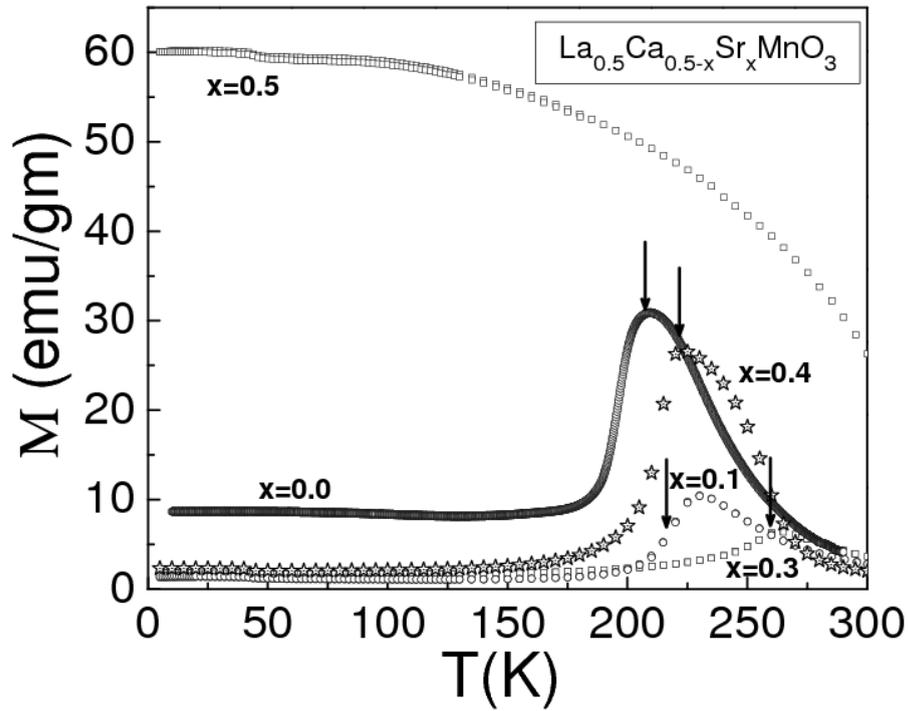

Figure: 4. The variation of magnetization with temperature in H = 1T for samples x = 0 - 0.5. For comparison the magnetization data for x =0.0 is taken from Ref. 7. The arrows in this figure show the endothermic peak transition temperature obtained from DSC.

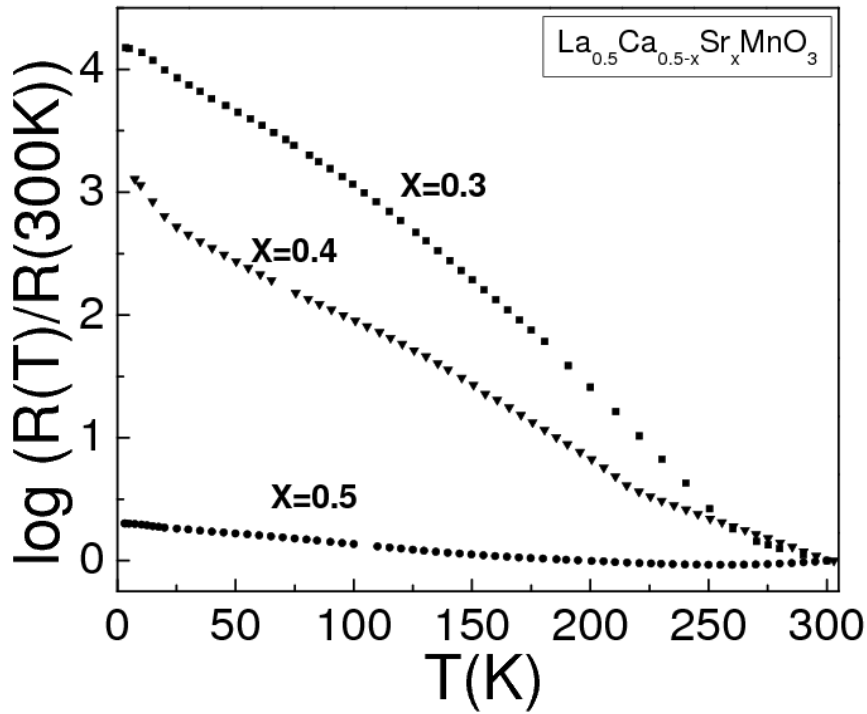

Figure: 5. The temperature dependence of normalized resistance for $La_{0.5}Ca_{0.5-x}Sr_xMnO_3$ (x = 0.3, 0.4, and 0.5) samples.

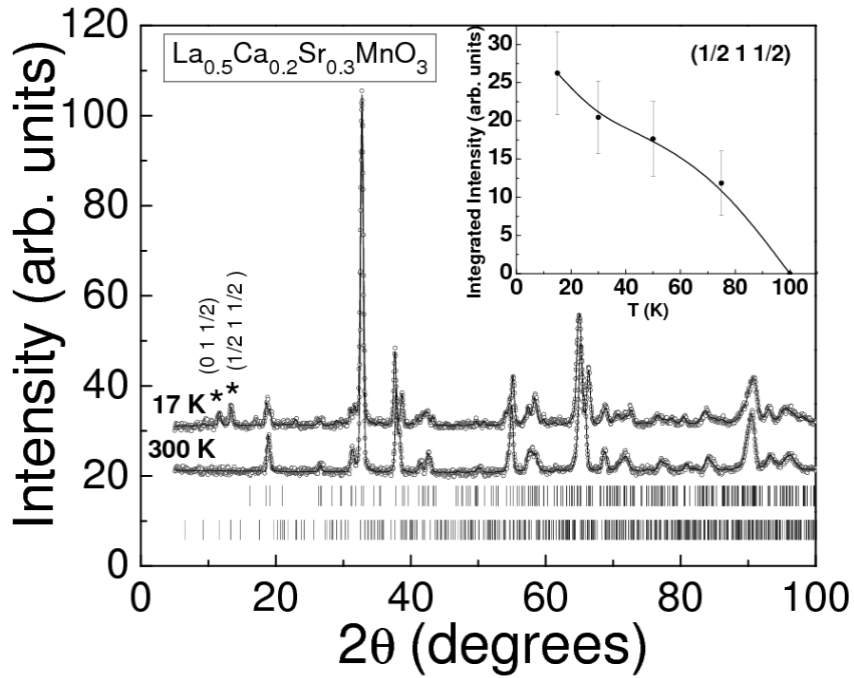

Figure: 6. Neutron diffraction pattern recorded on sample $La_{0.5}Ca_{0.2}Sr_{0.3}MnO_3$ at 17K and 300K. The symbol (*) indicates the AFM superlattice reflections. Continuous lines through the data points are the fitted lines to the chemical and magnetic structure described in the text. The inset shows the temperature variation of the antiferromagnetic superlattice reflection (½ 1 ½). The continuous lines are a guide for the eye.

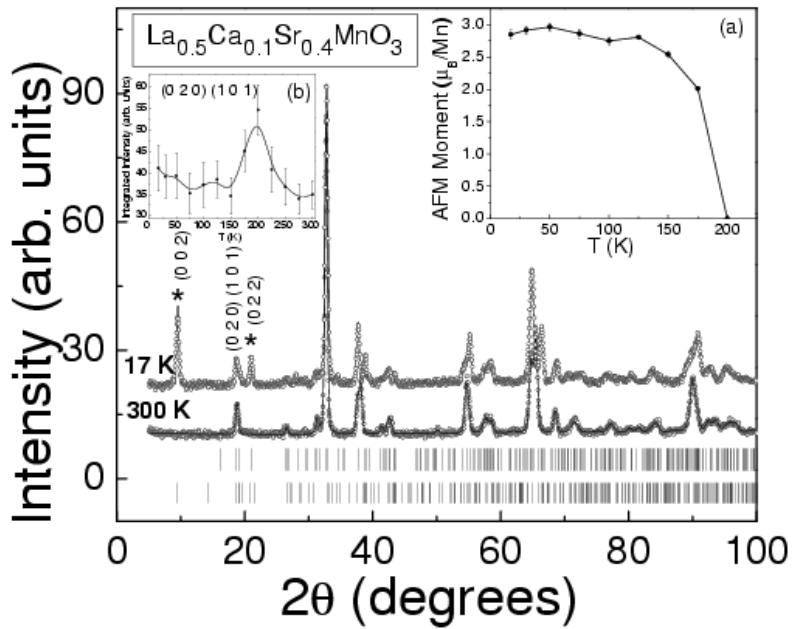

Figure: 7. Neutron diffraction pattern recorded on sample $La_{0.5}Ca_{0.1}Sr_{0.4}MnO_3$ at 17K and 300K. The symbol (*) indicates the antiferromagnetic reflections. Continuous lines through the data points are the fitted lines to the chemical and magnetic structure described in the text. The inset (a) shows the variation of antiferromagnetic moment with temperature. The inset (b) shows the temperature dependence of the ferromagnetic reflection (0 2 0). The continuous lines are a guide for the eye.

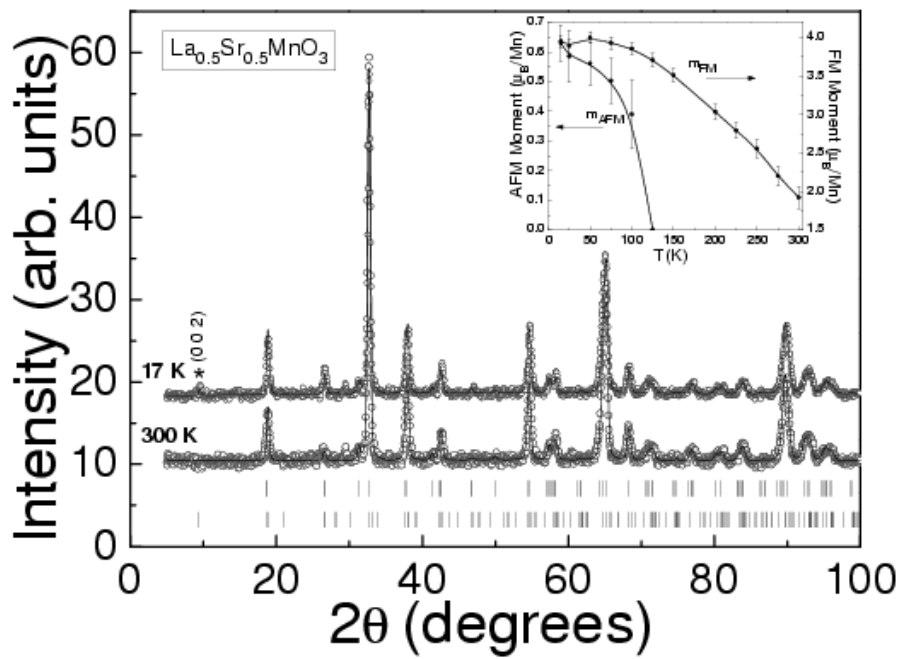

Figure: 8. Neutron diffraction pattern recorded on sample $La_{0.5}Sr_{0.5}MnO_3$ at 17K and 300K. The symbol (*) indicates the antiferromagnetic reflection. The inset shows temperature dependence of ferromagnetic and antiferromagnetic moment for $La_{0.5}Sr_{0.5}MnO_3$. The continuous lines are a guide for the eye.

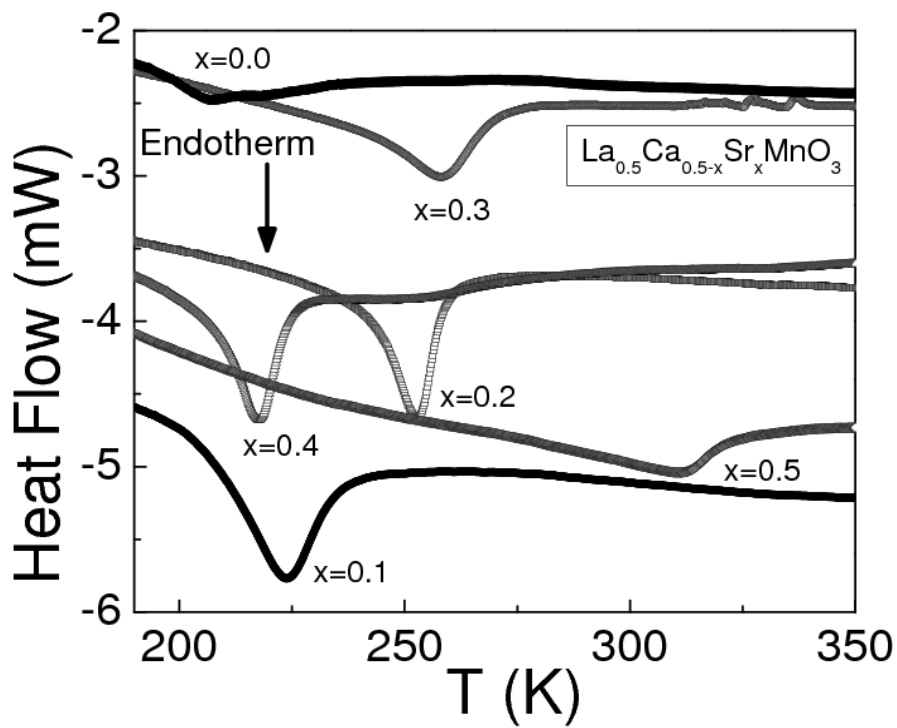

Figure: 9. The heat flow vs. temperature for $La_{0.5}Ca_{0.5-x}Sr_xMnO_3$ (x = 0.2 - 0.5) samples in the heating cycle.

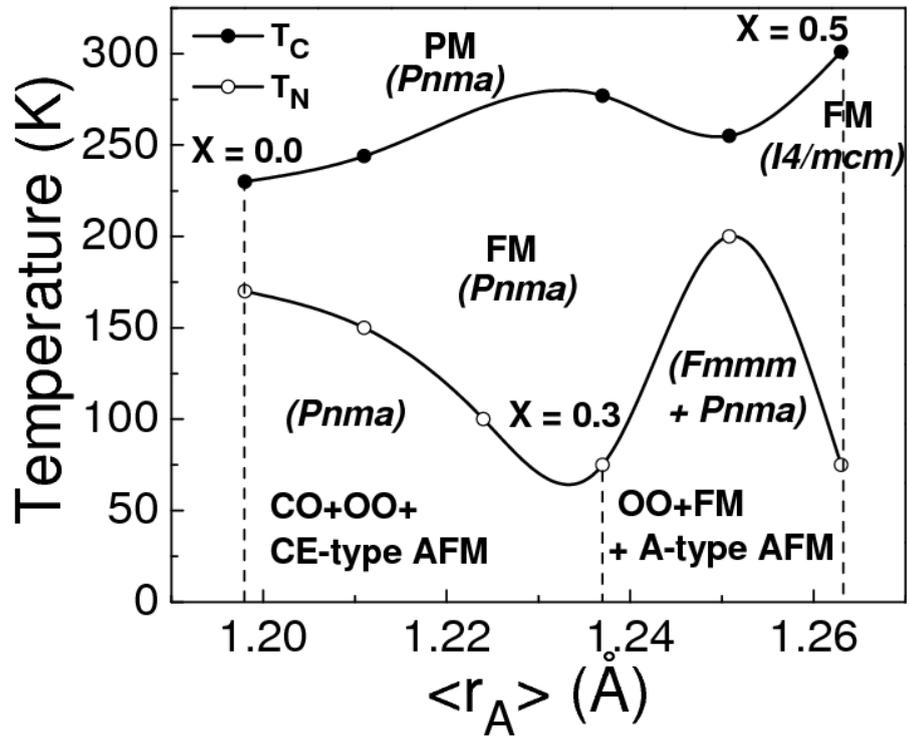

Figure: 10. The magnetic phase diagram of system $La_{0.5}Ca_{0.5-x}Sr_xMnO_3$. The space groups are indicated in brackets.

Table I. La$_{0.5}$Ca$_{0.5-x}$Sr$_x$MnO$_3$: Structural parameters obtained from Rietveld refinement of neutron diffraction pattern at 300K.

| Refined parameters | x = 0.1 *Pnma* | x = 0.2 *Pnma* | x = 0.3 *Pnma* | x = 0.4 *Pnma* | x = 0.4 *Fmmm* | x = 0.5 *I4/mcm* |
|---|---|---|---|---|---|---|
| a (Å) | 5.4249 (9) | 5.4263 (7) | 5.4326 (4) | 5.4435 (2) | 7.641 (2) | 5.4446 (2) |
| b (Å) | 7.646 (1) | 7.6189 (9) | 7.6133 (5) | 7.6347 (6) | 7.7632 (8) | 5.4446 (2) |
| c (Å) | 5.4519 (8) | 5.4525 (8) | 5.4662 (4) | 5.4721 (3) | 7.659 (2) | 7.7615 (7) |
| V (Å$^3$) | 226.14 (6) | 225.42 (5) | 226.08 (3) | 227.4 (2) | 454.3 (2) | 230.08 (3) |
| Mn-O$_1$ (Å) | 1.9356 (12) | 1.92607 (4) | 1.9248 (8) | 1.9255 (4) | 1.9224 (5) | 1.94037 (2) |
| Mn-O$_{21}$ (Å) | 1.951 (9) | 1.94687 (4) | 1.9504 (72) | 1.9136 (6) | 1.9261 (5) | 1.932 (30) |
| Mn-O$_{22}$ (Å) | 1.936 (9) | 1.92707 (4) | 1.9248 (73) | 1.9697 (7) | 1.9408 (2) | - |
| Mn-O$_1$-Mn (°) | 161.88 (5) | 162.887 (5) | 162.88 (4) | 164.81 (18) | 169.69 (7) | 180 |
| Mn-O$_2$-Mn (°) | 163.24 (4) | 166.261 (5) | 167.82 (31) | 167.24 (8) | 165.26 (7) | 169.99 (14) |
| Mn/ $\mu_B$ (FM) | - | - | - | 1.7 (2) | - | 3.95 (8) |
| $<r_A>$ (Å) | 1.211 | 1.224 | 1.237 | 1.25 | 1.25 | 1.263 |
| s$^2 \times 10^{-3}$ | 1.377 | 2.092 | 2.469 | 2.508 | 2.508 | 2.209 |
| T$_N$ (K) | 150 | 100 | 75 | 200 | 200 | 125 |
| (La, Sr, Ca) (x, y, z) | {0.015(6), ¼, -0.005(2)} | {0.006(3), ¼, -0.021 (1)} | {0.009 (2), ¼,-0.002 (1)} | {0.0005 (13), ¼, -0.5(7)} | {0, 0.242(2), 0} | {0, ½, ¼} |
| Mn (x, y, z) | {0, 0, ½} | {0, 0, ½} | {0, 0, ½} | {0, 0, ½} | | {0, 0, 0} |
| O (1) (x, y, z) | {0.494(3), ¼, 0.056(1)} | {0.507 (5), ¼, 0.0287 (7)} | {0.506 (3), ¼, ¼,} | {0.51 (2), ¼, 0.384(6)} | {0.226(2), 0, 0} | {0, 0, ¼} |
| O (2) (x, y, z) | {0.273(1), 0.0301(7), -0.271(2)} | {0.265 (2), 0.02283 (5), -0.266 (1)} | {0.258 (2), 0.0243 (3), -0.260 (1)} | {0.26(1), 0.200(4), -0.260(9)} | {0, 0, 0.281(3)} | {0.2281 (6), 0.7281 (6), 0} |
| B (Å$^2$) (La, Sr, Ca) | 0.40(6) | 0.28 (6) | 0.14 (4) | 0.3(4) | 0.6(1) | 0.62 (6) |
| B (Å$^2$) O (1) | 0.6(1) | 0.68 (9) | 0.58 (7) | 0.7(7) | 0.8(4) | 0.7 (2) |
| B (Å$^2$) O (2) | 0.81(8) | 0.47 (6) | 0.48 (5) | 0.7(4) | 0.7(3) | 0.59 (7) |
| B (Å$^2$) O (3) | - | - | - | | 1.3(2) | - |

Table II. La$_{0.5}$Ca$_{0.5-x}$Sr$_x$MnO$_3$: Structural parameters obtained from Rietveld refinement of neutron diffraction pattern at 17K.

| Refined parameters | x = 0.1 Pnma | x = 0.2 Pnma | x = 0.3 Pnma | x = 0.4 Pnma | x = 0.4 Fmmm | x = 0.5 I4/mcm |
|---|---|---|---|---|---|---|
| a (Å) | 5.4426(5) | 5.4366 (6) | 5.4406 (5) | 5.447 (1) | 7.5241 (2) | 5.4359 (2) |
| b (Å) | 7.5347(8) | 7.5237 (9) | 7.5289 (7) | 7.547 (2) | 7.763 (2) | 5.4359 (2) |
| c (Å) | 5.4884(6) | 5.4853 (7) | 5.4937 (6) | 5.498 (1) | 7.490(9) | 7.7529 (6) |
| V (Å$^3$) | 225.07(4) | 224.37 (5) | 225.03 (4) | 226.00 (9) | 451.2 (2) | 229.09 (2) |
| Mn-O$_1$ (Å) | 1.9156(11) | 1.9074 (2) | 1.9040 (2) | 1.9032 (4) | 1.9349 (4) | 1.9382 (2) |
| Mn-O$_{21}$ (Å) | 1.928(8) | 1.9176 (2) | 1.9109 (104) | 1.9316 (3) | 1.8934 (4) | 1.9310 (27) |
| Mn-O$_{22}$ (Å) | 1.978(3) | 1.9720 (2) | 1.9779 (100) | 1.9640 (3) | 1.9408 (5) | - |
| Mn-O$_1$-Mn (°) | 159.07(4) | 160.89 (3) | 162.65 (4) | 164.95 (6) | 172.98 (5) | 180 |
| Mn-O$_2$-Mn (°) | 163.4(3) | 166.23 (2) | 167.6 (4) | 166.71 (4) | 166.88 (5) | 168.88 (11) |
| Mn/µ$_B$ (AFM) | 1.5(1) | 1.39(4) | 1.30(4) | - | 2.85 (6) | 0.63(6) |
| (La, Sr, Ca) (x, y, z) | {0.0129(2), ¼, | {0.013(2), ¼, | {0.005(3), ¼, | {-0.009(6), ¼, | {0, 0.256 (2), 0} | {0, ½, ¼} |
| Mn (x, y, z) | {0, 0, ½} | {0, 0, ½} | {0, 0, ½} | {0, 0, ½} | {¼, 0, ¼ } | {0, 0, 0} |
| O(1) (x, y, z) | {0.492(3), ¼, | {0.508(3), ¼, | {0.507(4), ¼, | {0.507(8), ¼, 0.043(2)} | {0.227(4), 0, 0} | {0, 0, ¼} |
| O(2) (x, y, z) | {0.264(2), 0.0331(5), | {0.260(2), 0.0288(4), | {0.253(3), 0.0265(3), | {0.25(1), 0.027(1), | {0, 0, 0.279(2)} | {0.2257(5), 0.5257(5), 0} |
| O(3) (x, y, z) | - | - | - | - | {¼, ¼, ¼} | - |
| B(Å$^2$) (La, Sr, Ca) | 0.74(7) | 0.28 (8) | 0.19 (7) | 0.2 (2) | 0.2 (2) | 0.40 (5) |
| B (Å$^2$) O(1) | 0.71(9) | 0.40 (9) | 0.33 (7) | 0.6 (2) | 0.9 (4) | 0.6 (1) |
| B (Å$^2$) O(2) | 1.1(8) | 0.40 (6) | 0.48 (6) | 0.6(2) | 0.3(3) | 0.21 (5) |
| B (Å$^2$) O(3) | - | - | - | - | 1.2(4) | - |

Table III. Summary of transition temperature, and change in entropy obtained from DSC study during heating cycle.

| x | Entropy Change J/(K mole) | Transition Temperature (K) |
|---|---|---|
| 0.0 | 0.73 | 206.3 |
| 0.1 | 2.33 | 224 |
| 0.2 | 2.44 | 252.74 |
| 0.3 | 2.85 | 258.27 |
| 0.4 | 4.14 | 217.6 |
| 0.5 | 0.85 | 310.69 |